\documentclass[lettersize,journal]{IEEEtran}
\usepackage{amsmath,amsfonts}
\usepackage{array}
\usepackage[caption=false,font=normalsize,labelfont=sf,textfont=sf]{subfig}
\usepackage{textcomp}
\usepackage{stfloats}
\usepackage{url}
\usepackage{verbatim}
\usepackage{graphicx}
\usepackage{balance}
\usepackage{siunitx}
\usepackage{optidef}
\usepackage{makecell}
\usepackage{algorithm2e}
\RestyleAlgo{ruled}

\begin{document}
\title{Unsupervised Learning for Scalable Downlink Power Control in Cell-Free Massive MIMO}
\author{Giovanni Di Gennaro, Amedeo Buonanno, \IEEEmembership{Senior Member, IEEE}, Gianmarco Romano, Francesco Verde, \IEEEmembership{Senior Member, IEEE}, Stefano Buzzi, \IEEEmembership{Fellow, IEEE}, and Francesco A.N.~Palmieri, \IEEEmembership{Member, IEEE}
\thanks{
G.~Di Gennaro was supported by the Italian Ministry for University and Research (MUR) - PON Ricerca e Innovazione 2014–2020 (D.M. 1062/2021). 
F.A.N.~Palmieri was also supported by POR CAMPANIA FESR 2014/2020, A-MOBILITY: Technologies for Autonomous Vehicles.}
\thanks{G.~Di Gennaro, G.~Romano, F.~Verde, and F.A.N.~Palmieri are with Dipartimento di Ingegneria, Università degli Studi della Campania ``Luigi~Vanvitelli'', Aversa (CE), 81031, Italy [e-mails:~(giovanni.digennaro; gianmarco.romano; francesco.verde; francesco.palmieri)@unicampania.it].
A. Buonanno is with the Department of Energy Technologies and Renewable Sources, ENEA, Portici (NA), 80055, Italy (e-mail: amedeo.buonanno@enea.it).
S. Buzzi is with University of Cassino and Southern Lazio, I-03043 Cassino, Italy, and with Consorzio Nazionale Interuniversitario per le Telecomunicazioni, I-43124 Parma, 
Italy (e-mail: buzzi@unicas.it).}
\thanks{Digital Object Identifier xxxxxxxxxxxxxxxxxxx.}
}
\markboth{IEEE Wireless Communications Letters,~Vol.~xx,
No.~yy,~Month~2026}{Di Gennaro \MakeLowercase{\textit{et al.}}:
Physics-Informed Unsupervised Learning for Scalable Downlink Power Control in Cell-Free Massive MIMO}

\IEEEpubid{0000--0000/00\$00.00~\copyright~2026 IEEE}

\maketitle

\begin{abstract}
In cell-free massive multiple-input multiple-output systems, downlink power control is essential to ensure uniformly high service quality across users.
Existing solutions range from computationally demanding centralized optimization to scalable distributed schemes with limited fairness.
This letter proposes an unsupervised, physics-informed framework that directly optimizes max-min fairness without optimal labels or user position information.
The method scales with the number of UEs, requires no retraining when the number of UEs changes, and admits a fully scalable extension.
Numerical results show that it nearly doubles the worst-user spectral efficiency over existing scalable schemes.
\end{abstract}


\section{Introduction}

\IEEEPARstart{C}{ell}-free massive multiple-input multiple-output (CF-mMIMO)
has emerged as a promising architecture for beyond-5G networks, where
distributed access points (APs) jointly serve the user equipments (UEs)
without cell boundaries~\cite{Ngo2017}. Its distributed architecture
provides macro-diversity and robustness against shadowing and blockages \cite{cellfreemagazine}.

The large number of APs and UEs, however, places a substantial burden on
fronthaul and central processing units (CPUs), motivating scalable
designs~\cite{Bjornson2020}. Downlink power control is particularly
challenging: centralized approaches targeting max-min fairness (MMF) or
sum-rate maximization~\cite{Ngo2017,buzzi2019user} require global channel
knowledge and iterative optimization, whereas scalable distributed
strategies such as fractional power control~\cite{Interdonato2019,lozano2020}
rely on local information but provide limited worst-user spectral-efficiency
gains.

This letter proposes a scalable power-control policy based on a
bidirectional long short-term memory (BiLSTM) network. The closest related
work \cite{Chafaa2025} employs a Transformer for MMF power control, but
relies on centralized processing, UE positions, and optimal supervision
labels, without guaranteeing per-AP power constraints. In contrast, our
physics-informed unsupervised policy uses only large-scale fading
information, enforces per-AP constraints by construction, and accommodates
different numbers of UEs without retraining.

\IEEEpubidadjcol

\section{System model and baseline methods}
\label{sec:cell-free}
We consider the downlink of a CF-mMIMO system with $K$ single-antenna UEs and $L$ distributed APs, each with $M$ antennas. 
The system operates in time-division duplex (TDD), where uplink and downlink share the same frequency band over alternating time intervals. 
Under the assumption of channel reciprocity, downlink channel estimates are obtained from uplink pilots. 
Each coherence block spans $\tau_c$ symbols, partitioned into $\tau_p$, $\tau_u$, and $\tau_d$ symbols for uplink pilots, uplink data, and downlink data, respectively, with $\tau_c = \tau_p + \tau_u + \tau_d$.
We model channels between AP $\ell \in \mathcal{L} = \{1, \dots, L\}$ and UE $k \in \mathcal{K} = \{1, \dots, K\}$ as
$\mathbf{h}_{k,\ell} = \sqrt{\beta_{k,\ell}} \, \tilde{\mathbf{h}}_{k,\ell} \in \mathbb{C}^M$,
where $\tilde{\mathbf{h}}_{k,\ell} \sim \mathcal{N}_\mathbb{C}(\mathbf{0}_M, \mathbb{I}_M)$ 
captures small-scale fading and $\beta_{k,\ell}$ accounts for large-scale fading (LSF) \cite{Demir2021}.

To initiate communication, each UE selects a \emph{master AP} based on the strongest large-scale channel gain, which subsequently coordinates pilot assignment and data transmission.
Once a pilot is assigned, the master AP informs the network (via the CPU), enabling all APs to determine their suitability for serving the UE. 
Based on this information, each UE $k$ is associated with a subset $\mathcal{L}_k \subset \mathcal{L}$ of APs for data transmission, with AP-UE association typically determined either via heuristic methods \cite{Demir2021} or optimized procedures \cite{digennaroAPUE}.
Under linear minimum mean square error (MMSE) channel estimation and normalized maximum-ratio (MR) beamforming, with $\mathbf{w}_{k,\ell}=\widehat{\mathbf{h}}_{k,\ell}/\sqrt{M\gamma_{k,\ell}}$ and $\mathbb{E}\{\|\mathbf{w}_{k,\ell}\|^2\}=1$, the achievable spectral efficiency (SE) is lower-bounded \cite{Demir2021} as
\begin{equation}
	\mathrm{SE}_k = \frac{\tau_d}{\tau_c} \log_2 \left(1 + \mathrm{SINR}_k\right)  \qquad {\textrm{[bit/s/Hz]}}
\label{eq:SE}
\end{equation}
where 
\begin{equation}
    \mathrm{SINR}_k = \resizebox{0.73\hsize}{!}{$\displaystyle
    \frac{\displaystyle M \!\left(\sum_{\ell \in \mathcal{L}_k} \!\!\sqrt{\rho_{k,\ell} \gamma_{k,\ell}} \right)^{\!\!2}}{ \displaystyle \sum_{i \in \mathcal{K}} \sum_{\ell \in \mathcal{L}_i} \rho_{i,\ell}\beta_{k,\ell} + M\!\!\!\sum_{i \in \mathcal{P}_k \setminus\{k\}}\!\!\left(\sum_{\ell \in \mathcal{L}_i} \!\!\sqrt{\rho_{i,\ell} \gamma_{k,\ell}} \right)^{\!\!2}\! + \sigma_\mathrm{dl}^2}$}
    \label{eq:SINRk}
\end{equation}
with $\rho_{k,\ell}\geq 0$ denoting the power allocated by AP $\ell$ to UE $k$. 

\noindent
The parameter $\gamma_{k,\ell}$ is given by
\begin{equation}
    \gamma_{k,\ell} = \frac{\tau_p \eta_k \beta_{k,\ell}^2}{\displaystyle\tau_p\!\!\sum_{i \in \mathcal{P}_{k}}\!\!\eta_i \beta_{i,\ell} + \sigma_{\mathrm{ul}}^2}
\end{equation}
where $\eta_k$ denotes  the transmit power of UE $k$ during uplink channel estimation, $\sigma_{\mathrm{ul}}^2$ is the noise variance in the uplink, and $\mathcal{P}_{k}$ represents the set of UEs that share the same pilot with UE $k$.
The numerator in \eqref{eq:SINRk} represents the desired-signal power, where the contributions from the APs in $\mathcal L_k$ combine coherently and $\gamma_{k,\ell}$ accounts for the channel-estimation quality. The denominator consists of the non-coherent multiuser interference, which depends on the total transmit power radiated by each AP weighted by the corresponding LSF coefficient $\beta_{k,\ell}$, the coherent interference generated by UEs sharing the pilot of UE $k$, which scales with $M$ similarly to the desired signal, and the downlink noise power $\sigma_\mathrm{dl}^2$.

As a primary baseline, we consider the distributed fractional power
allocation (FPA) scheme in~\cite{Interdonato2019}, which allocates the transmit 
power from AP $\ell$ to UE $k$ according to
\begin{equation}
    \rho_{k,\ell} = P_{\ell}^\mathrm{max} \frac{(\beta_{k,\ell})^\nu}{\displaystyle\sum_{i \in \mathcal{K}_\ell} (\beta_{i,\ell})^\nu}
    \label{eq:FPL}
\end{equation}
where $P_{\ell}^\mathrm{max}$ denotes the power budget available at AP $\ell$, $\mathcal{K}_\ell$ is the set of UEs served by AP $\ell$, and $\nu \in [-1,1]$ is a design parameter. 
Equation \eqref{eq:FPL} can be computed locally at AP $\ell$ using only the LSF coefficients between AP $\ell$ and the UEs in $\mathcal{K}_\ell$, which ensures full scalability of the scheme.

A second baseline, proposed in \cite{lozano2020}, assigns transmit power according to
\begin{equation}
    \rho_{k,\ell} \propto 
    \frac{\beta_{k,\ell}}
    {\displaystyle \left(\sum_{j \in \mathcal{L}} \beta_{k,j}\right)^{\!\!\theta}
     \left(\sum_{i \in \mathcal{K}} \frac{\beta_{i,\ell}}{\left(\sum_{j \in \mathcal{L}} \beta_{i,j}\right)^{\!\theta}}\right)^{\!\!\zeta}}
    \label{eq:baseline_dl_fpc}
\end{equation}
where $\theta \in [0, 1]$ and $\zeta \in [0.4, 1.6]$.
Equation \eqref{eq:baseline_dl_fpc} refers to a cell-free scenario where all the APs in the system serve all the UEs in the system (i.e., ${\cal L}_k={\cal L}$).  
After enforcing the per-AP power constraint via normalization, the dependence on $\zeta$ vanishes and \eqref{eq:baseline_dl_fpc} reduces to 
\begin{equation}
    \rho_{k,\ell}= P_{\ell}^\mathrm{max} \frac{\beta_{k,\ell} }
    {\displaystyle \left(\sum_{j \in \mathcal{L}} \beta_{k,j}\right)^{\!\!\theta} 
    \sum_{i \in \mathcal{K}} \frac{\beta_{i,\ell}}{(\sum_{j \in \mathcal{L}} \beta_{i,j})^{\theta}}} \, .
\end{equation}
As an additional baseline, we include equal power allocation (EPA), in which  AP $\ell$ equally distributes its available power among its served UEs, i.e., $\rho_{k,\ell} = P_\ell^{\mathrm{max}}/|\mathcal{K}_\ell|$ for all $k \in \mathcal{K}_\ell$.

\section{Deep Learning Approach}
\label{sec:dl}
We introduce the proposed unsupervised downlink power control algorithm.
After the AP--UE association has been determined, the serving sets $\mathcal{L}_1,\ldots,\mathcal{L}_K$ are known. 
Given these associations, the remaining quantities to be determined are the downlink power coefficients $\rho_{k,\ell}$ assigned by AP $\ell$ to UE $k$. 
Although the proposed framework can be trained for various differentiable objective functions (e.g., sum-SE maximization or weighted multi-objective utilities), in this letter we consider the maximization of the minimum user spectral efficiency, since the weakest UE determines the perceived service quality in user-centric CF-mMIMO. 
The design problem is therefore
\begin{maxi!}
    {\scriptstyle\{\rho_{k,\ell}\}_{\ell\in\mathcal{L},\,k\in\mathcal{K}_\ell}}
    {\min_{k \in \mathcal{K}} \mathrm{SE}_k\!\left( \{\rho_{i,\ell}\}_{\substack{\ell\in\mathcal{L},\\ i\in\mathcal{K}_\ell}}\mid \mathcal{L}_1, \ldots, \mathcal{L}_K\right) \label{eq:objectiveProblem}}
    {\label{eq:Problem}}{}
    \addConstraint{\rho_{k,\ell}}{\geq 0 \quad}{\forall \ell \in \mathcal{L},\; \forall k \in \mathcal{K}_\ell \label{eq:C2Problem}}
    \addConstraint{\sum_{i \in \mathcal{K}_\ell} \rho_{i,\ell}}{\leq P_{\ell}^\mathrm{max} \quad}{\forall \ell \in \mathcal{L} \:. \label{eq:C1Problem}}
\end{maxi!}
Problem~\eqref{eq:Problem} couples the AP-UE coefficients through both the
interference terms in \eqref{eq:SINRk} and the per-AP power constraints \eqref{eq:C1Problem},
precluding independent link optimization.
It can be solved via iterative MMF optimization,
providing a performance upper bound. However, its centralized
implementation requires global topology and channel statistics and must be recomputed whenever they change, resulting in poor scalability.

We replace online optimization with a learned parametric policy mapping channel descriptors to feasible power coefficients, i.e., $\{\rho_{k,\ell}\}_{\ell,k} = \pi_{\Theta}\!\left(\{\beta_{k,\ell}\}_{k,\ell}\right)$, which is trained offline.
The training is unsupervised and physics-informed: the loss is computed from the communication utility in \eqref{eq:SE}--\eqref{eq:SINRk}, not from labels generated by a separate optimizer, thus avoiding the costly label-generation stage.
Feasibility is imposed directly by the output layer and by an AP-wise normalization, so that \eqref{eq:C2Problem}--\eqref{eq:C1Problem} are satisfied by construction.

\begin{figure}[!t]
    \centering
    \includegraphics[width=0.9\linewidth]{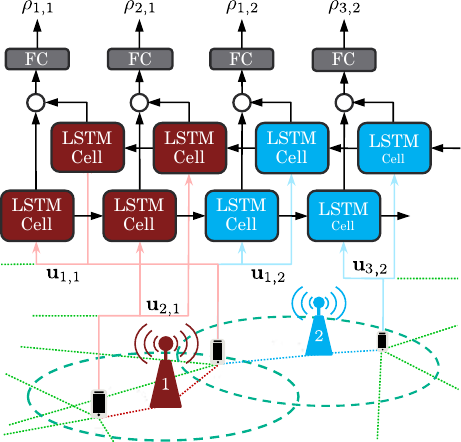}
    \caption{Contextual schematic of the AP-centric input construction. For each AP $\ell$, the served UEs are represented through the corresponding UE-dependent quantities, arranged into the input sequence processed by the BiLSTM.}
    \label{fig:bilstm_context}
\end{figure}

The neural architecture is AP-centric, as shown in Fig.~\ref{fig:bilstm_context}. 
For AP $\ell$, the network processes only the UEs in $\mathcal{K}_\ell$ and produces the coefficients $\{\rho_{k,\ell}:k\in\mathcal{K}_\ell\}$. 
The same parameters are reused across all APs and sequence positions, which allows the same model to handle APs with different numbers of served UEs without architectural modifications.

As a BiLSTM operates on ordered sequences, a fixed deterministic ordering is assigned to the APs to ensure consistency across evaluations \cite{digennaroAPUE}. 
During training, the UEs within each subset are instead randomly permuted, preserving the AP-level structure while reducing the reliance on specific UE positions. 
This results in a hierarchical AP–UE representation in which the AP-specific UE subsets form the higher-level sequence processed by the BiLSTM. 
The bidirectional architecture allows each AP representation to incorporate contextual information from the other APs, which is relevant given the coupling among power-control decisions in \eqref{eq:SINRk} and \eqref{eq:C1Problem}. 
Moreover, the shared recurrent parameters provide a lightweight and parameter-efficient architecture capable of handling APs serving different numbers of UEs. 
Compared with conventional fully connected networks, the proposed recurrent architecture can accommodate variations in the number of UEs served by each AP without modifying the network architecture or retraining the model. 
Graph neural networks provide a natural alternative for modeling the AP–UE interactions, but require explicit graph construction and message-passing operations, whereas Transformer-based architectures generally incur higher computational and memory costs due to their attention mechanism. 
Overall, the BiLSTM offers a suitable trade-off between modeling the AP–UE interactions, supporting variable-sized serving sets without retraining, and maintaining computational and memory efficiency.

For each active pair $(k,\ell)$, the input feature vector is
\begin{equation}
\mathbf{u}_{k,\ell} = \Big[ \beta_{k,\ell},\;
\sum_{j\in\mathcal{L}}\beta_{k,j},\;
\sum_{i\in\mathcal{K}}\beta_{i,\ell} \Big]^\top
\label{eq:features_vector}
\end{equation}
where the entries represent the direct AP--UE coupling, the aggregate large-scale channel strength observed by UE $k$, and the aggregate coupling between AP $\ell$ and the UE population. 
These quantities have fixed dimension for any $K$ and $L$, and they do not require UE-position information. 
In the implementation, they are expressed on a base-10 logarithmic scale, since this improves the numerical conditioning of the input features when the propagation losses are heterogeneous. 
The required LSF coefficients coincide with the statistics already exploited for uplink pilots, pilot assignment, and AP-UE association.

Let $\{\mathbf{u}_{t,\ell}\}_{t=1}^{|\mathcal{K}_\ell|}$ denote the input feature sequence at AP $\ell$, with $\mathbf{u}_{t,\ell}=\mathbf{u}_{k_t,\ell}$ and $k_t$ denoting the UE placed at position $t$ in the sequence. 
The BiLSTM computes the forward and backward hidden representations according to
\begin{equation}
\begin{aligned}
\overrightarrow{\mathbf{s}}_{t,\ell} &= \mathrm{LSTM}_{\mathrm{f}}\!\left(\mathbf{u}_{t,\ell},\overrightarrow{\mathbf{s}}_{t-1,\ell}\right) \\
\overleftarrow{\mathbf{s}}_{t,\ell} &= \mathrm{LSTM}_{\mathrm{b}}\!\left(\mathbf{u}_{t,\ell},\overleftarrow{\mathbf{s}}_{t+1,\ell}\right)
\end{aligned}
\label{eq:bilstm_rec}
\end{equation}
and obtains the representation by element-wise summation
\begin{equation}
\mathbf{s}_{t,\ell}=\overrightarrow{\mathbf{s}}_{t,\ell}+\overleftarrow{\mathbf{s}}_{t,\ell}
\label{eq:context_sum}
\end{equation}
which acts as the aggregation function of the BiLSTM layer. 
Hence, $\mathbf{s}_{t,\ell}$ embeds the information of UE $k_t$ together with the contextual interactions induced by the other UEs.
Before entering the prediction head, the BiLSTM output is mapped back to the linear domain as
\begin{equation}
\widetilde{\mathbf{s}}_{t,\ell}
=
10^{\mathbf{s}_{t,\ell}}.
\label{eq:linear_mapping_bilstm}
\end{equation}
The transformed representation $\widetilde{\mathbf{s}}_{t,\ell}$ is then processed by a stack of fully connected layers with progressively decreasing dimensionality. 
The hidden dense layers employ SELU activation functions, while the final output layer adopts a Softplus activation to guarantee strictly positive power coefficients. 
Accordingly, the predicted coefficient is expressed as
\begin{equation}
\widehat{\rho}_{k,\ell}
=
\log\!\left(
1+
e^{y_{k,\ell}}
\right)
>0
\label{eq:softplus_output}
\end{equation}
where $y_{k,\ell}$ denotes the pre-activation output of the final layer.

To enforce the per-AP transmit power constraint, we define
\begin{equation}
S_\ell
=
\sum_{k\in\mathcal{K}_\ell}
\widehat{\rho}_{k,\ell}
\quad \text{and} \quad
\alpha_\ell
=
\min\!\left\{
1,
\frac{P_\ell^{\mathrm{max}}}{S_\ell}
\right\}
\label{eq:ap_scaling_factor}
\end{equation}
and the transmitted power coefficients become
\begin{equation}
\rho_{k,\ell}
=
\alpha_\ell
\widehat{\rho}_{k,\ell},
\qquad
\forall \ell\in\mathcal{L},
\ \forall k\in\mathcal{K}_\ell.
\label{eq:projected_power}
\end{equation}
This normalization is independently applied at each AP and guarantees the satisfaction of constraints \eqref{eq:C2Problem} and \eqref{eq:C1Problem}.

The trainable parameters $\Theta$ are learned by minimizing a differentiable soft-min surrogate of the MMF utility over mini-batches of independent network realizations.
For a mini-batch of size $B$, the resulting loss function is defined as
\begin{equation}
\mathcal{L}(\Theta) = \frac{1}{TB} \sum_{b=1}^{B} \log\!\left[\sum_{k\in\mathcal{K}} e^{-T\,\mathrm{SE}^{(b)}_k 
\left(\{\rho^{(b)}_{k,\ell}(\Theta)\}_{k,\ell}\right)}\right]
\label{eq:training_loss_softmin}
\end{equation}
where the temperature $T>0$ controls the sharpness of the approximation, and each realization comprises the LSF coefficients and AP–UE association of a single network snapshot.
The SE values are evaluated after the coefficients have been generated by \eqref{eq:bilstm_rec}--\eqref{eq:projected_power}, so the gradients propagate through the complete computation graph from the loss to the network parameters.
Therefore, unlike supervised learning, no power-control labels are required, as the loss is computed from the achievable SE in (1)–(2) resulting from the generated coefficients. Consequently, no MMF bisection, online feasibility test, or label-generation stage is required.

\begin{algorithm}[!t]
\caption{Unsupervised training procedure}\label{alg:training}
\KwData{LSF and AP–UE associations per snapshot}
\KwResult{Learned parameters $\Theta^{\star}$}

Initialize network parameters $\Theta$\;
\For{$e \gets 1$ \KwTo $N_{\mathrm{epochs}}$}{
    Shuffle training set and construct mini-batches $\mathcal{B}$\;
    \ForEach{mini-batch $\mathcal{B}$}{
        Construct AP-centric input features $\mathbf{u}_{k,\ell}$ for all served pairs $(k,\ell)$ \;
        Compute BiLSTM contextual representations and latent outputs $y_{k,\ell}$ \;
        Map latent outputs to transmit powers and apply AP-wise normalization using \eqref{eq:ap_scaling_factor}--\eqref{eq:projected_power}\;
        Evaluate $\mathcal{L}(\Theta)$ from the resulting SE values \eqref{eq:SE}--\eqref{eq:SINRk} through 
        \eqref{eq:training_loss_softmin}\;
        Backpropagate gradients and update $\Theta$\;
    }
}
\end{algorithm}

Algorithm~\ref{alg:training} summarizes the resulting training procedure.
Note that the SINR and SE expressions are needed only during this offline stage, as they define the physics-informed unsupervised training objective.
At inference time, AP $\ell$ forms the local sequence for the UEs in $\mathcal{K}_\ell$, evaluates the shared BiLSTM-based policy, and applies 
\eqref{eq:ap_scaling_factor}--\eqref{eq:projected_power}. 
Moreover, the same trained model can be applied to a different number of UEs, since the recurrent policy accepts variable-length sequences and shares its parameters across positions.

\section{Scalability and complexity analysis}
\label{sec:scalability}
Scalability refers to the behavior of a system as the network size increases while UE and AP densities remain fixed.
In this regime, the deployment area grows, and consequently also the total number of APs and UEs increases. A scheme is scalable if the computational complexity, memory requirements, and information exchange per network element remain finite and do not grow with the overall network size.

Under this assumption, the proposed method is scalable with respect to the number of UEs, since each AP performs inference only over its associated UEs, 
whose cardinality is bounded by the user-centric association mechanism and by the local network density.
However, strict per-AP scalability cannot rely on input features that aggregate information over the entire sets of APs and UEs.
Therefore, in order to make the proposed architecture fully scalable, the feature vector in \eqref{eq:features_vector} must be constructed by replacing the sums over the full sets $\mathcal{L}$ and $\mathcal{K}$ with sums over local neighborhoods, similar to what is done  in \cite{digennaroAPUE}. 
More precisely, the aggregate channel strength of UE $k$ should be computed only over the APs comprised in an area (large but finite) surrounding UE $k$, while the aggregate load/coupling observed by AP $\ell$ should be computed only over the UEs deployed in an area (large but finite) surrounding AP $\ell$. 
Accordingly, a scalable version of \eqref{eq:features_vector} can be written as
\begin{equation}
\mathbf{u}_{k,\ell} =
\left[
\beta_{k,\ell},
\sum_{j\in\mathcal{N}_k^{\rm AP}}\beta_{k,j},
\sum_{i\in\mathcal{N}_\ell^{\rm UE}}\beta_{i,\ell}
\right]^\top
\label{eq:features_vector_scalable}
\end{equation}
where $\mathcal{N}_k^{\rm AP}$ denotes the set of APs in the neighborhood of UE $k$, and $\mathcal{N}_\ell^{\rm UE}$ denotes the set of UEs in the neighborhood of AP $\ell$. 
These neighborhoods can be defined, for instance, according to distance, large-scale fading thresholds, or the user-centric association rule. 
If the network density is kept fixed, the cardinalities of $\mathcal{N}_k^{\rm AP}$ and $\mathcal{N}_\ell^{\rm UE}$ remain bounded as the deployment area increases, and hence the dimension and construction cost of the input descriptors remain finite.

In user-centric CF-mMIMO systems, the online complexity depends on the active AP-UE associations, i.e., on the number of UEs served by each AP, $|\mathcal{K}_{\ell}|$. 
As such, a general complexity characterization is nontrivial, since it depends on both the global association topology and the user distribution across APs, even under association strategies that limit the per-AP load. 
However, in the reference scenario considered herein, each UE is associated with a fixed number $N$ of APs, yielding exactly $NK$ active connections. 
Therefore, it is straightforward to see that the overall inference complexity of the proposed network scales linearly with the number of UEs, i.e., $\mathcal{O}(K)$.
This is a substantial improvement over bisection-based MMF algorithms, whose complexity is dominated by the repeated solution of a linear feasibility problem and scales as $\mathcal{O}(K^3)$ per bisection step \cite{buzzi2019user}.
For each active pair $(k,\ell)$, the BiLSTM processes the input vector $\mathbf{u}_{k,\ell} \in \mathbb{R}^{d_{\text{in}}}$ in \eqref{eq:features_vector}, where $d_{\text{in}}=3$.
Assuming one MAC equals two FLOPs and including both directions plus hidden-state summation, the BiLSTM cost is $2 \cdot 8 H (H + d_{\text{in}} + 1) + H \approx 1.06$ MFLOPs for $H=256$.
The prediction head ($256 \to 64 \to 16 \to 1$) contributes $\approx 0.04$ MFLOPs per pair, resulting in a total inference cost of approximately $1.10$ MFLOPs per evaluation.
With $N=4$ active pairs per UE, the total inference cost scales as $4 \cdot K \cdot 1.10 \approx 4.4 \cdot K$ MFLOPs.
We further note that the policy depends only on large-scale statistics and must therefore be re-evaluated only when these, or the AP-UE association derived from them, change, rather than on a per-coherence-block basis.

From a memory perspective, all computations are performed in FP32 precision (4 bytes per parameter), and the model architecture is fixed and independent of the number of active pairs $(k,\ell)$ or timesteps.
The BiLSTM with $H=256$ contains $532,480$ parameters, while the prediction head contributes $17,505$ parameters, resulting in a total of $549,985$ parameters. 
This yields a constant memory footprint of $\approx 2.10$ MiB.

\section{Simulation Results}
\label{sec:simulations}
Numerical simulations are conducted in a square area of $500\times 500\,\mathrm{m}^2$, with $K=8$ UEs, $L=16$ APs, and \mbox{$M=4$} antennas per AP. 
We assume a coherence block length of $\tau_c = 200$ symbols and, for all considered values of $K$, assign orthogonal pilot sequences by setting $\tau_p = K$, thereby avoiding pilot contamination.\footnote{Increasing $\tau_p$ with $K$ reduces the pre-log factor ${\tau_d}/{\tau_c}$ in \eqref{eq:SE}, introducing a spectral efficiency penalty that is common to all evaluated schemes.}
This simplifying assumption can be relaxed via a pilot reuse strategy as in \cite{digennaroAPUE}.
The uplink transmit power for each UE is $100\,\mathrm{mW}$, whereas the maximum downlink transmit power for each AP is $200\,\mathrm{mW}$. 
LSF coefficients are computed according to the 3GPP microcell model~\cite[Table B.1.2.1-1]{36.814} for a $2\,\mathrm{GHz}$ carrier frequency, a pathloss exponent of $3.67$, and a UE-AP height difference of $10\,\mathrm{m}$. 
Shadow fading is modeled as $F_{k\ell}\sim\mathcal{N}(0,\sigma_F^2)$
with $\sigma_F=4\,\mathrm{dB}$, and shadowing terms are spatially correlated to account for similar blocking conditions experienced by closely located UEs~\cite{36.814}. 
The noise power is $\sigma^2=-94\,\mathrm{dB}$, the noise figure is $7\,\mathrm{dB}$, and the system bandwidth is $20\,\mathrm{MHz}$.
As standard in user-centric CF-mMIMO, AP–UE association is performed prior to power control. 
Since the AP–UE association strategy is not the focus of this work, we fix the association size to $N=4$, with each UE being associated with the four APs having the largest LSF coefficients.

The BiLSTM is trained with mini-batches of size 64, 256 hidden units in the BiLSTM module, 
the SGD optimizer with momentum $0.9$, and a learning rate of $10^{-2}$. 
The prediction head is implemented with two fully connected hidden layers with $64$ and $16$ units, followed by a scalar output layer that provides the latent coefficient $\widehat{\rho}_{k,\ell}$ used in \eqref{eq:softplus_output} for power mapping.

The soft-min temperature is set to $T=10$, which bounds the gap between the surrogate in \eqref{eq:training_loss_softmin} and the true minimum SE by $\ln(K)/T$, while retaining informative gradients across UEs.
The remaining hyperparameters of the proposed architecture are determined through a preliminary tuning phase on a representative validation set, \emph{disjoint from the test set}. 
A limited range of candidate values is considered for the main architectural and training parameters, and the configuration providing the best trade-off among communication performance, convergence stability, and computational complexity is selected. 
The resulting hyperparameters are then kept fixed throughout training and evaluation, without adaptation to individual network realizations.

Training is performed on a fixed set of $8000$ UE positions, which are shuffled and partitioned into $1000$ groups of $8$ UE at each epoch, yielding $1000$ randomly formed network snapshots per epoch.
Testing is performed on 200 groups of 8 UE-position realizations, disjoint from the training set. 

\begin{figure}[!t]
    \centering
    \includegraphics[width=\linewidth]{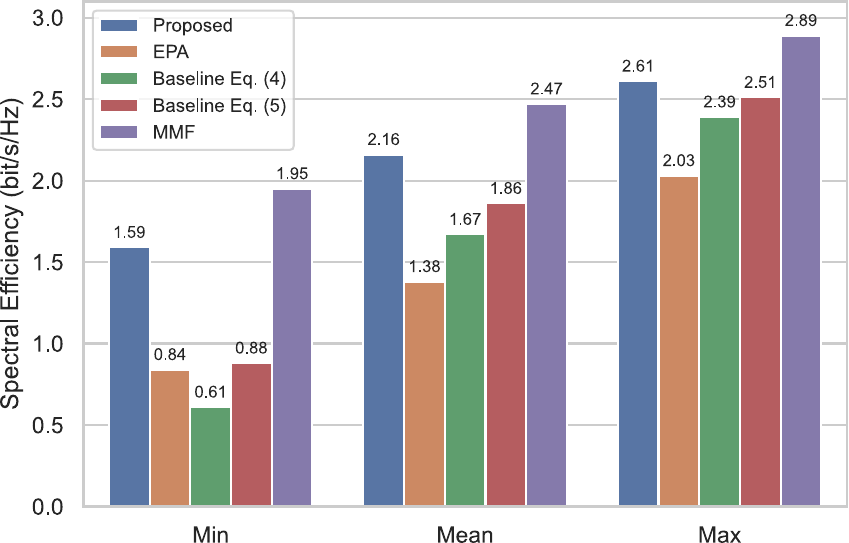}
    \vspace{-15pt}
    \caption{Average test performance (bit/s/Hz).}
    \label{fig:table}
\end{figure}

As shown in Fig.~\ref{fig:table}, compared to the considered scalable baselines, the proposed method nearly doubles the minimum SE on the test set, while also yielding improvements in both average and maximum SE.
Although a gap remains with respect to the centralized MMF upper bound, the proposed method achieves this performance without requiring global channel knowledge, iterative optimization, or retraining as the network topology changes.
A direct comparison with \cite{Chafaa2025} is not included because it considers $K$ per-UE powers under a network-wide sum-power constraint, whereas our user-centric formulation optimizes per-AP–UE powers under separate per-AP constraints in \eqref{eq:C1Problem}. 
Unlike the closed-form rule in \eqref{eq:baseline_dl_fpc}, which admits direct per-AP normalization, adapting \cite{Chafaa2025} would require redefining the network outputs and regenerating the supervision labels.

\begin{figure}[!t]
    \centering
    \includegraphics[width=\linewidth]{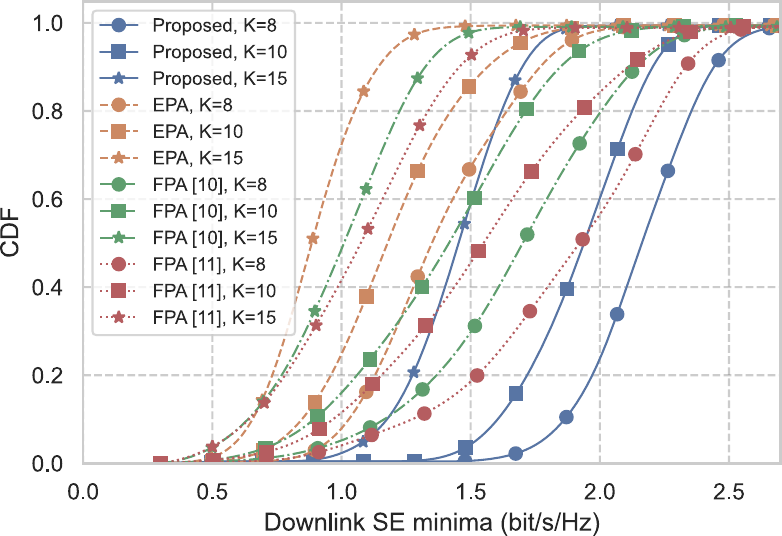}
    \caption{CDF of the SE minima, where a model trained with $K=8$ UEs is evaluated at $K=8$,  $K=10$,  and  $K=15$.}
    \label{fig:minima}
\end{figure}

Fig.~\ref{fig:minima} reports the empirical CDF of the per-realization minimum SE over the test set. 
The proposed approach consistently outperforms all scalable baselines, with the performance gain increasing with the number of users. 
Since the area is kept fixed, this corresponds to an increasing UE density and per-AP load.  
Notably, the learned policy maintains its performance advantage without retraining as the operating conditions move beyond those encountered during training, demonstrating its ability to generalize to previously unseen user densities and per-AP loads, in addition to its structural scalability.

\section{Conclusions}
\label{sec:conclusion}
We have proposed an unsupervised, physics-informed downlink power control framework for CF-mMIMO systems. 
The AP-centric BiLSTM policy enforces per-AP power constraints by construction, scales linearly with the number of UEs, and nearly doubles the worst-user spectral efficiency compared to scalable baselines, without requiring supervision labels, UE position information, or retraining when the user population changes.
Future work will validate the fully scalable extension, alternative fairness objectives, and pilot reuse.

\bibliographystyle{ieeetr}
\bibliography{IEEEabrv,bibliography}

\end{document}